\documentclass{mg11}

\newcommand{\be}{\begin{equation}}
\newcommand{\ee}{\end{equation}}
\newcommand{\bea}{\begin{eqnarray}}
\newcommand{\eea}{\end{eqnarray}}

\begin{document}



\title{ON ENERGY AND MOMENTUM OF THE FRIEDMAN AND SOME MORE GENERAL UNIVERSES}

\author{JANUSZ GARECKI}

\address{Institute of Physics, University of Szczecin,\\
Wielkopolska 15, 70-451 Szczecin, Poland\\
\email{garecki@sus.univ.szczecin.pl}}


\begin{abstract}
Recently some authors  concluded that the energy and momentum of the Fiedman universes, flat and closed,
are equal  to zero   locally and globally (flat universes) or only globally (closed universes).
The similar conclusion was also done for more general only homogeneous universes
(Kasner and Bianchi type I).
Such conclusions originated from coordinate dependent calculations performed only in comoving
Cartesian coordinates by using the so-called {\it energy-momentum complexes}.
By using new coordinate  independent expressions on energy and momentum
one can show that the Friedman and more general universes {\it needn't be energetic nonentity}.
\end{abstract}

\bodymatter

\noindent
In the last years many authors have calculated the energy and momentum of
the Friedman universes and also more general, only spatially
homogeneous universes, like Kasner, Bianchi type I and Bianchi
type II universes \cite{one}.

The above mentioned authors performed their calculations in
special comoving coordinates called ``Cartesian coordinates''
despite that they used {\it coordinate dependent} double index
energy-momentum complexes, matter and gravitation.
The all energy-momentum complexes {\it are neither geometrical
objects nor coordinate independent objects}, e.g., they can vanish
in some coordinates locally or globally and in other coordinates
they can be different from zero. It results that the double index
energy-momentum complexes and the gravitational energy-momentum
pseudotensors determined by them {\it have no physical meaning to
a local analysis} of a gravitational field, e.g., to study
gravitational energy distribution. In fact, up to now, complexes and pseudotensors
were reasonably used only to calculate the global quantities for the very
precisely defined asymptotically flat spacetimes (in spatial or in
null direction).
The best one of the all possible double index energy-momentum
complexes from physical and geometrical points of view is the
Einstein canonical double index energy momentum complex $_E K
_i^{~k}$ (See, e.g., \cite{Tr62,Gold80}).

The conclusion of the authors which calculated the energy and
momentum of the Friedman and more general universes by using
double index energy-momentum complexes is the following: the
energy and momentum of the closed Friedman universes {\it are equal to
zero globally}, and in the case of the flat Friedman universes and
their generalizations (Kasner, Bianchi type I, Bianchi type II universes)
these quantities {\it are equal to zero locally and globally}.

One can have at least the following objections against the
calculations of such a kind and against the above conclusion:
\begin{enumerate}
\item The authors despite that they used coordinate dependent
expressions had performed their calculations only in Cartesian
comoving coordinates.

The results obtained in other comoving coordinates, e.g., in
coordinates $(t,\chi,\vartheta,\varphi)$ or in coordinates $(t,r,\vartheta,\varphi)$
{\it are dramatically different}.
\item The local ``energy-momentum distribution'' as given by any
energy-momentum complex {\it has no physical sense} but the
authors try to give a physical sense   of this distribution, e.g.,
they assert that the total energy density for flat Friedman
universes, for Kasner and Bianchi type I universes, is null.
\item The conclusion leads us to Big-Bang which {\it has no
singularity} in total energy density.
\item The global energy and momentum {\it have physical meaning} only
when spacetime is asymptotically flat either in spatial or null
direction and when these quantities can be measured. But this is not a
case of the cosmological models.

So, the problem of the global energy and global linear (or angular) momentum
for Friedman, and for more general universes also, {\it is not
well-posed from the physical point of view} because these
universes are not asymptotically flat spacetimes, and, in
consequence, their global quantities {\it cannot be measurable}.
This problem can only have {\it a mathematical sense}.
\end{enumerate}
 Thus, one can doubt in physical validity  of the conclusion that the energy
 and momentum of the Friedman, Kasner, Bianchi type I and Bianchi
 type II universes {\it are equal to zero}; especially that all these
 universes {\it are energy-free}.

By using double index energy-momentum complexes one should rather
conclude that the energy and momentum of the Friedman, Kasner,
Bianchi type I, and Bianchi type II universes explicite depend on
the used coordinates and, therefore, they are undetermined {\it not only
locally} but also {\it globally}. The last conclusion is very
sensible because, as we mentioned beforehand, one cannot measure the global
energy and global linear (or angular) momentum of the Friedman and
any more general universe. One can do this only in the case of an
isolated system when spacetime is asymptotically flat.

One cannot use the coordinate independent {\it Pirani} and {\it
Komar} \cite{Tr62,Gold80}
expressions in order to correctly prove (at least from the
mathematical point of view) the statement that the energy of the
Friedman, Kasner, Bianchi type I and Bianchi type II universes
disappears, i.e., that these universes have zero net energy. It is
because we have no translational timelike Killing vector field
(descriptor of energy in Komar expression) in these universes, and
the privileged normal congruence of the fundamental observers
which exists in these universes is geodesic (Pirani
expression on energy only can be applied  in a spacetime having a
privileged normal and timelike congruence. But for a geodesic
congruence Pirani expression fails giving trivially zero).

One also cannot use for this purpose the coordinate independent
Katz-Bi\v cak-Lynden (BKL)  bimetric approach \cite{BKL96} because
the results obtained in this approach depend on the used
background and on mapping of the spacetime under study onto this
background.

Thus, the ``academic'' statement that the Friedman, Kasner,
Bianchi type I and Bianchi type II universes have no energetic
content {\it is still not satisfactory proved}. But by using Komar
expression, one can correctly (at least from mathematical point of view)
prove that the linear momentum for these universes disappears in a
comoving coordinates.

Recently we have introduced the new, coordinate independent expressions
on the averaged relative energy-momentum and angular momentum in general relativity
(See \cite{Gar8}). We have called these new tensorial expressions
{\it the averaged tensors of the relative energy-momentum and angular momentum}.
The averaged tensors are very closey related to the canonical superenergy
and angular supermomentum tensors which were introduced in our previous papers \cite{Mal1}.
When applied, the averaged relative energy-momentum tensors give the {\it positive-definite}
energy densities for the Friedman, Kasner and Bianchi type I  universes \cite{Gar8}.
The result of such a kind is very satisfactory from the physical point of view.
The more general universes were not analyzed yet.

\vfill

\end{document}